\documentclass[12pt,a4paper]{article}

\usepackage[T1]{fontenc}
\usepackage[utf8]{inputenc}
\usepackage{amsmath,amssymb,amsfonts}
\usepackage{graphicx}
\usepackage{bm}
\usepackage{slashed}
\usepackage{booktabs}
\usepackage{array}
\usepackage{hyperref}
\usepackage{geometry}
\usepackage{microtype}
\usepackage{needspace}
\usepackage{placeins}
\usepackage{xcolor}
\usepackage{titlesec}

\hypersetup{
  hidelinks,
  pdftitle={Scalar and spin-two energy-momentum-tensor structure in near-threshold charmonium probes of the proton},
  pdfauthor={Arkadiy I. Syamtomov},
  pdfsubject={Energy-momentum tensor structure in near-threshold charmonium probes},
  pdfkeywords={energy-momentum tensor, trace anomaly, charmonium, light-front, gravitational form factors, near-threshold photoproduction}
}
\titleformat{\section}
  {\normalfont\normalsize\bfseries}
  {\thesection}{0.7em}{}

\newcommand{\as}{a_s}
\newcommand{\MSbar}{\ensuremath{\overline{\text{MS}}}}
\newcommand{\calA}{\mathcal A}

\newcommand{\GeV}{\mathrm{GeV}}
\newcommand{\fms}{\mathrm{fm}}

\begin{document}

\title{Scalar and spin-two energy-momentum-tensor structure in near-threshold\\ charmonium probes of the proton}

\author{Arkadiy I.~Syamtomov\thanks{Corresponding author: \href{mailto:arkady.syamtomov@gmail.com}{\texttt{arkady.syamtomov@gmail.com}}}}
\date{\small\textit{Bogolyubov Institute for Theoretical Physics,\\
National Academy of Sciences of Ukraine,\\
Kyiv, Ukraine}}
\maketitle

\begin{abstract}
I analyze the energy-momentum-tensor (EMT) content of the chromoelectric operator that governs the leading interaction of compact charmonium with soft gluonic fields.  In the renormalized \MSbar{} operator basis this interaction separates into a scalar anomaly/sigma branch and a traceless spin-two branch.  The forward light-front spin-two term is fixed by partonic plus-momentum fractions and gives a controlled correction to scalar dominance.  Off forward, however, the chromoelectric spin-two contraction is not governed by $A_i(t)$ alone; it also contains the combination $3B_i(t)-D_i(t)$, equivalently $6J_i(t)-3A_i(t)-D_i(t)$.  This identifies which EMT structures charmonium can access in forward and differential observables, without interpreting the measured slope as a model-independent static three-dimensional mass radius.
\end{abstract}

\noindent\textit{Keywords:} energy-momentum tensor, trace anomaly, charmonium, light-front decomposition, gravitational form factors, near-threshold photoproduction

\section{Introduction}

Heavy quarkonium has long been used as a compact probe of gluonic fields in hadrons.  In the short-distance treatment of Peskin and Bhanot--Peskin, the small size of a heavy $Q\bar Q$ state allows its interaction with soft hadronic matter to be organized through a local operator expansion, with the leading coupling controlled by chromoelectric fields~\cite{Peskin1979I,BhanotPeskin1979II}.  Recent near-threshold $J/\psi$ measurements have brought this idea into the gravitational-form-factor (GFF) program, where differential photoproduction data are used to constrain gluonic EMT structure, scalar trace profiles, and mechanical distributions of the proton~\cite{Kharzeev2021MassRadius,Duran2023GluonicGFF,Jpsi007_2026,CLAS12_2026,GuoJiYuan2023,GuoYuanZhao2025,PentchevChudakov2024}.  The broader theoretical setting of the nucleon EMT and its GFFs is reviewed in Refs.~\cite{BurkertRMP2023,PolyakovSchweitzer2018}.

The aim of this work is to determine how the established charmonium chromoelectric interaction filters the scalar and spin-two components of the nucleon EMT.  The existence of a rich quark/gluon trace and EMT decomposition is well known, in particular from the renormalized trace analysis of Hatta, Rajan and Tanaka~\cite{HattaRajanTanaka2018} and Tanaka's higher-order extension~\cite{Tanaka2019Trace}.  The question addressed here is more specific: once the chromoelectric operator is inserted between proton states, which scalar anomaly/sigma and traceless spin-two matrix elements does it select, what is their relative size in the forward light-front limit, and how do the corresponding GFFs enter the measured off-forward $t$ dependence?

This operator viewpoint is useful because the chromoelectric probe does not project onto a single scalar gluonic EMT observable.  The scalar branch controls the dominant forward anomaly/sigma normalization, while the spin-two branch is tied to light-front momentum flow and, off forward, to the gravitational structures that describe angular momentum and mechanical response.  In particular, the differential cross section can receive shape modifications from $B_i(t)$ and $D_i(t)$ through the same chromoelectric contraction that reduces to the familiar plus-momentum contribution at $t=0$.

The light-front formulation also clarifies the interpretation of slopes.  Recent discussions emphasize that nucleon radii inferred from local-operator matrix elements should not automatically be read as literal three-dimensional boundaries of the proton~\cite{Petrov2024Radii,Petrov2026Sizes}.  In the $\xi=0$ light-front frame, however, $t=-\Delta_\perp^2$, and form-factor slopes can be interpreted as transverse impact-parameter profile scales when the density construction is applicable~\cite{FreeseMiller2021,FreeseMiller2021Unified}.  I therefore use the language of scalar and gravitational transverse slopes, or light-front transverse profile scales, rather than static three-dimensional mass radii.

The paper is organized as follows.  Section~\ref{sec:operator} gives the chromoelectric operator and the scalar/spin-two EMT basis.  Section~\ref{sec:forward} evaluates the forward light-front matrix element and defines the spin-two/scalar diagnostic ratio.  Section~\ref{sec:offforward} derives the full off-forward contraction and identifies the $3B_i-D_i$ combination.  Section~\ref{sec:numerics} summarizes the benchmark and robustness analysis.  Appendices collect convention checks, coefficient evolution, the off-forward derivation, and the dimensional matching derivation of Eq.~\eqref{eq:OE_match}.

\section{Chromoelectric operator and EMT basis}\label{sec:operator}

I start from the heavy-quark short-distance picture of Peskin and Bhanot--Peskin, in which a compact quarkonium state couples to soft gluonic fields through a color-dipole, or chromoelectric, operator expansion~\cite{Peskin1979I,BhanotPeskin1979II}.  In the notation used below, the local operator selected by the leading chromoelectric interaction takes the form
\begin{equation}
  \mathcal O_E(v)=g^2\mathbf E^2(v)
  =-g^2\left(F^{a\mu\alpha}v_\alpha\right)
          \left(F^{a}_{\ \mu\beta}v^\beta\right),
\label{eq:OE_def}
\end{equation}
where $v^\mu$ is the quarkonium four-velocity.  The quarkonium polarizability and process-dependent normalization multiply the matrix element of Eq.~\eqref{eq:OE_def}; they are not needed for the EMT separation itself.

The mostly-minus metric is used, $g^{\mu\nu}=\mathrm{diag}(1,-1,-1,-1)$.  The gluon spin-two operator is
\begin{equation}
  O_g^{\mu\nu}
  =-F^{a\mu\alpha}F^{a\nu}_{\ \ \alpha}
  +\frac{1}{4}g^{\mu\nu}F^2,
\label{eq:Og_def}
\end{equation}
with an analogous singlet quark operator $O_q^{\mu\nu}$.  The renormalized quark/gluon separation of the trace is a scheme-defined operator question.  Hatta, Rajan and Tanaka showed in dimensional regularization with minimal subtraction how the quark and gluon contributions to the QCD trace anomaly are associated with the renormalized quark and gluon EMTs, and how this structure constrains the twist-four GFFs $\bar C_{q,g}$~\cite{HattaRajanTanaka2018}; Tanaka extended the formula to three loops~\cite{Tanaka2019Trace}.  I use this renormalized EMT framework, but organize the scalar part of the chromoelectric matching in the $(\calA,S)$ basis introduced below.  This choice avoids assigning a unique observable meaning to a separate quark or gluon trace contribution.  Such quark/gluon trace partitions are scheme conventions, often discussed in the D1/D2 and $(x,y)$ languages of the proton mass-decomposition literature~\cite{MetzPasquiniRodini2020,LorceMetzPasquiniRodini2021}.  The scalar matrix element used here is therefore the anomaly/sigma basis appropriate to the matched chromoelectric operator, not a standalone identification of a D1 gluon trace with an observable mass component.

In the renormalized \MSbar{} operator basis, the chromoelectric operator is
\begin{equation}
[g^2\mathbf E^2(v)]_{\MSbar}
=
 g^2 v_\mu v_\nu O_g^{\mu\nu}
-\frac{g^2v^2}{4}[F^2]_R
+\frac{g^2v^2}{4}\gamma_m S
+O(\as^2) .
\label{eq:OE_match}
\end{equation}
Here $S=\sum_q m_q\bar q q$, $\gamma_m$ is the quark-mass anomalous dimension, and $\as=\alpha_s/(4\pi)$.  Equation~\eqref{eq:OE_match} is the basic operator result used in the rest of the paper.  It is a renormalized one-loop \MSbar{} operator identity: the trace subtraction is performed in $d=4-2\epsilon$ before minimal subtraction, so the finite $\gamma_m S$ term is the remnant of the evanescent spin-two projector rather than an independent phenomenological addition.  The derivation is given in Appendix~\ref{app:matching}.  Thus the chromoelectric probe contains a traceless spin-two branch and a scalar branch.  The scalar branch is most transparent in the basis
\begin{equation}
  \calA = \frac{\beta(g)}{2g}[F^2]_R+\gamma_m S,
  \qquad S=\sum_qm_q\bar q q .
\label{eq:A_basis}
\end{equation}
With covariant normalization of proton states,
\begin{equation}
  \calA_N\equiv\frac{1}{2M}\langle N|\calA|N\rangle=M-\Sigma_N,
  \qquad
  \Sigma_N=\sum_q\sigma_{qN},
\label{eq:A_sigma_matrix}
\end{equation}
where $\sigma_{qN}=(2M)^{-1}\langle N|m_q\bar q q|N\rangle$.  Thus the scalar contribution is the anomaly complement of the sigma terms together with the sigma matrix element.  It is not the beta-function operator alone at finite quark mass.

The main operator structures used below are summarized in Table~\ref{tab:operator_meaning}.  The table is included here because the same symbols reappear in the forward light-front ratio and in the off-forward GFF contraction.

\begin{table}[!htbp]
\centering
\caption{Operator structures and physical roles used in the chromoelectric EMT decomposition.}
\label{tab:operator_meaning}
\begin{tabular}{lll}
\toprule
Structure & Matrix element or definition & Physical role \\
\midrule
$\calA$ & $(2M)^{-1}\langle N|\calA|N\rangle=M-\Sigma_N$ & anomaly complement to sigma terms\\
$S$ & $(2M)^{-1}\langle N|S|N\rangle=\Sigma_N$ & scalar sigma contribution\\
$A_i(t)$ & second moment of $H_i$ at $\xi=0$ & LF momentum/spin-two form factor\\
$B_i(t)$ & second moment of $E_i$ at $\xi=0$ & gravitomagnetic/spin-flip form factor\\
$J_i(t)$ & $\frac12[A_i(t)+B_i(t)]$ & angular-momentum GFF\\
$D_i(t)$ & mechanical GFF & pressure/shear contribution\\
\bottomrule
\end{tabular}
\end{table}
\FloatBarrier

In the scalar basis $(\calA,S)$ the reduced forward scalar amplitude is
\begin{equation}
  \frac{N_0(0)}{M}
  = C_{\calA}\left(1-\frac{\Sigma_N}{M}\right)
    + C_\sigma\frac{\Sigma_N}{M} .
\label{eq:N0_forward}
\end{equation}
At one-loop matching for $v^2=1$ and $n_f=3$,
\begin{equation}
  C_{\calA}=\frac{8\pi^2}{\beta_0},
  \qquad
  C_\sigma=\gamma_m\left(\frac{g^2}{4}-\frac{8\pi^2}{\beta_0}\right),
  \qquad \beta_0=11-\frac{2}{3}n_f .
\label{eq:scalar_coeffs}
\end{equation}

The active-flavor convention is fixed as follows.  For proton scalar matrix elements I use the low-energy $n_f=3$ basis
\begin{equation}
   \calA^{(3)}=\Theta-\sum_{q=u,d,s}m_q\bar q q .
\label{eq:nf3_basis}
\end{equation}
If the hard matching scale is chosen above a heavy-quark threshold, the coefficient functions are first evolved in the corresponding active-flavor theory and then matched onto the low-energy scalar basis by standard decoupling.  At leading order in the heavy-quark expansion,
\begin{equation}
   \calA^{(n_l)} = \calA^{(n_l+1)} + m_Q\bar Q Q +O(1/m_Q^2),
\end{equation}
with the heavy-quark scalar matrix element related to the light-flavor anomaly matrix element by the familiar heavy-quark theorem~\cite{ShifmanVainshteinZakharov1978}.  The numerical analysis below keeps the scalar proton matrix element in the $n_f=3$ basis; the matching-scale variation of the spin-two coefficients is then treated separately.

Having separated the scalar and spin-two operators, the next question is how these structures are evaluated in proton states.  The scalar matrix element is fixed by Eq.~\eqref{eq:A_sigma_matrix}; the spin-two part requires a clean interpretation of $v_\mu v_\nu O_i^{\mu\nu}$.  This motivates the light-front analysis of the next section, where the trace-free plus component selects partonic momentum flow.

\section{Forward light-front interpretation}\label{sec:forward}

The light-front formulation is introduced here because it gives a direct physical meaning to the forward spin-two matrix element.  In ordinary covariant notation the contraction $v_\mu v_\nu O_i^{\mu\nu}$ mixes several components of the traceless EMT.  On the light front, by contrast, the plus-plus component contains no trace term at all and its forward matrix element is the partonic plus-momentum fraction.

Light-front coordinates are defined as $x^\pm=(x^0\pm x^3)/\sqrt2$, with $g^{+-}=1$ and $g^{++}=0$.  The sectoral EMT split
\begin{equation}
  T_i^{\mu\nu}=O_i^{\mu\nu}+\frac14 g^{\mu\nu}\theta_i
\end{equation}
implies
\begin{equation}
  T_i^{++}=O_i^{++}.
\label{eq:Tpp_tracefree}
\end{equation}
Thus the plus-plus component is trace-free and selects the second moment of the partonic light-front momentum distribution.  In the forward limit,
\begin{equation}
  A_g(\mu)=\int_0^1 dx\,xg(x,\mu),
  \qquad
  A_q(\mu)=\sum_q\int_0^1dx\,x[q(x,\mu)+\bar q(x,\mu)],
\end{equation}
with $A_q+A_g=1$ for the total EMT.

For threshold kinematics with $v$ aligned with the proton momentum, the reduced spin-two matrix element is
\begin{equation}
  \frac{1}{2M}\langle P|v_\mu v_\nu O_i^{\mu\nu}|P\rangle_{\rm thr}
  =\frac34 M A_i .
\label{eq:three_fourths}
\end{equation}
This is a spin-two light-front momentum-flow contribution.  It is not a scalar trace or mass contribution.

I denote by $C_i^{(2)}(\mu,\mu_h)$ the Wilson coefficients multiplying the traceless spin-two quark and gluon EMT operators after matching at the hard scale $\mu_h$ and evolution to the light-front scale $\mu$.  At the matching scale, the leading chromoelectric operator is gluonic,
\begin{equation}
  C_g^{(2)}(\mu_h)=g^2(\mu_h),
  \qquad
  C_q^{(2)}(\mu_h)=0,
\end{equation}
while singlet spin-two evolution generates $C_q^{(2)}\ne0$ when $\mu\ne\mu_h$.  The explicit evolution is given in Appendix~\ref{app:rg}.

The forward matrix element therefore contains two distinct amplitudes: a scalar amplitude controlled by Eq.~\eqref{eq:N0_forward}, and a traceless spin-two amplitude controlled by the same Wilson coefficients that multiply the PDF momentum fractions.  To quantify whether the spin-two term is leading, negligible, or a controlled correction to the scalar amplitude, I define
\begin{equation}
R_{2/0}^{\rm LF}(0)=
\frac{\frac34\left[C_q^{(2)}A_q+C_g^{(2)}A_g\right]}
{C_{\calA}(1-f_\sigma)+C_\sigma f_\sigma},
\qquad f_\sigma=\frac{\Sigma_N}{M}.
\label{eq:R_forward}
\end{equation}
The numerator and denominator in Eq.~\eqref{eq:R_forward} are normalized by the same factor of $M$.  The numerical ratios use $M=0.9383~\GeV$, $\Sigma_N=59~\mathrm{MeV}$, hence $f_\sigma=0.0629$, and $\alpha_s^{(3)}(2~\GeV)=0.3066$ for the low-energy scalar coefficients.  Varying $\Sigma_N$ over $45$--$75~\mathrm{MeV}$ changes the central CT18NNLO $\mu_h=2~\GeV$ value by about $-1.9\%$ to $+2.2\%$, smaller than the displayed matching-scale scan.

Table~\ref{tab:forward_pdf} gives the PDF-based forward inputs at $Q_{\rm LF}=2~\GeV$.  The LHAPDF-grid values were treated as $x f(x,Q)$, so that $A_g=\int_0^1 dx\,xg(x,Q)$ is obtained directly by integrating the tabulated gluon entry.  ABMP16\_5 was not used at $Q=2~\GeV$ because its grid begins at a higher scale.

\begin{table}[!htbp]
\centering
\caption{Forward PDF moments and single-scale spin-two/scalar ratio at $Q_{\rm LF}=2~\GeV$.  The quark singlet moment $A_q^{\rm act}$ is the active-flavor singlet returned by the PDF grid; the small charm component $A_c$ is shown separately as a flavor check.  The scale-separated scan is shown in Table~\ref{tab:scale_scan}.}
\label{tab:forward_pdf}
\begin{tabular}{lccccc}
\toprule
PDF set & $A_g$ & $A_{uds}$ & $A_c$ & $A_g+A_q^{\rm act}$ & $R_{2/0}^{\rm LF}(\mu_h=2~\GeV)$\\
\midrule
CT18NNLO & 0.4134 & 0.5760 & 0.0106 & 1.0000 & 0.1470\\
MSHT20nnlo & 0.4080 & 0.5835 & 0.0087 & 1.0002 & 0.1449\\
NNPDF4.0 NNLO & 0.3994 & 0.5886 & 0.0119 & 1.0000 & 0.1419\\
\bottomrule
\end{tabular}
\end{table}

The scalar matrix element in Eq.~\eqref{eq:nf3_basis} is kept in the low-energy $n_f=3$ anomaly/sigma basis.  The traceless spin-two numerator, however, is evaluated in the active-flavor PDF scheme of each grid at $Q_{\rm LF}=2~\GeV$.  I verified the flavor content explicitly: the charm momentum fraction in the grids is only $A_c=0.0087$--$0.0119$.  Repeating the scale scan with only the $u,d,s$ quark moment leaves the $\mu_h=2~\GeV$ entries unchanged, because $C_q^{(2)}(\mu_h)=0$, and changes the largest-$\mu_h$ entries by less than $0.5\%$.  Thus the quoted $R_{2/0}^{\rm LF}$ values should be read as active-flavor spin-two moments matched to an $n_f=3$ scalar denominator, with negligible numerical ambiguity from the small charm momentum at this scale.

The spin-two fraction is stable across modern PDF sets.  The larger variation comes from the matching scale, because the spin-two coefficient decreases as the hard scale is raised.  This is summarized in Table~\ref{tab:scale_scan}.

\begin{table}[!htbp]
\centering
\caption{Scale dependence of $R_{2/0}^{\rm LF}(0)$ for the three PDF inputs.}
\label{tab:scale_scan}
\begin{tabular}{lcccc}
\toprule
PDF set & $\mu_h=2$ GeV & $\mu_h=3$ GeV & $\mu_h=5$ GeV & $\mu_h=10$ GeV\\
\midrule
CT18NNLO & 0.1470 & 0.1303 & 0.1144 & 0.0996\\
MSHT20nnlo & 0.1449 & 0.1288 & 0.1133 & 0.0987\\
NNPDF4.0 NNLO & 0.1419 & 0.1266 & 0.1115 & 0.0974\\
\bottomrule
\end{tabular}
\end{table}

\FloatBarrier

The forward conclusion is that the scalar anomaly/sigma amplitude dominates the normalization, while the light-front spin-two contribution is a controlled correction.  The remaining question is whether this hierarchy survives once the momentum transfer is resolved.  This leads to the off-forward analysis, where the same chromoelectric contraction is expressed in terms of gravitational form factors and becomes sensitive to $B_i(t)$ and $D_i(t)$.

\section{Off-forward contraction and gravitational structure}\label{sec:offforward}

The forward analysis fixes the normalization of the scalar and spin-two contributions at $t=0$, but charmonium photoproduction is measured through differential cross sections.  The observed $t$ dependence therefore depends on the off-forward continuation of the same chromoelectric operator.  In this section I derive the off-forward spin-two contraction and show that it probes not only the momentum form factor $A_i(t)$, but also the gravitomagnetic and mechanical form factors $B_i(t)$ and $D_i(t)$.

The off-forward matrix element of a sectoral EMT is conventionally written as~\cite{Ji1997,BurkertRMP2023,HackettPefkouShanahan2024}
\begin{align}
\langle P'|T_i^{\mu\nu}|P\rangle
=\bar u(P')\bigg[
&A_i(t)\gamma^{(\mu}\bar P^{\nu)}
+B_i(t)\frac{\bar P^{(\mu}i\sigma^{\nu)\alpha}\Delta_\alpha}{2M}
\nonumber\\
&+D_i(t)\frac{\Delta^\mu\Delta^\nu-g^{\mu\nu}\Delta^2}{4M}
+\bar C_i(t)M g^{\mu\nu}
\bigg]u(P),
\label{eq:EMT_GFF}
\end{align}
where $\bar P=(P'+P)/2$, $\Delta=P'-P$, $t=\Delta^2$, and $\bar P^2=M^2-t/4$.  The convention in Eq.~\eqref{eq:EMT_GFF} uses $D_i(t)/(4M)$; if a source uses a coefficient $C_i(t)$ multiplying the same tensor divided by $M$, then $D_i(t)=4C_i(t)$.

In the symmetric light-front frame with $\Delta^+=0$, equivalently $\xi=0$, the plus component remains trace-free because $g^{++}=0$.  However, the chromoelectric operator contains the full contraction $v_\mu v_\nu O_i^{\mu\nu}$, not only $O_i^{++}$.  With $v^\mu=\bar P^\mu/\sqrt{\bar P^2}$, the reduced helicity-conserving contraction is
\begin{equation}
\frac{1}{2M}\langle P'|v_\mu v_\nu O_i^{\mu\nu}|P\rangle
=
\sqrt{1-\frac{t}{4M^2}}
\left[\frac34MA_i(t)
+\frac{t}{16M}\bigl(3B_i(t)-D_i(t)\bigr)\right].
\label{eq:full_contraction_ABD}
\end{equation}
The form factor
\begin{equation}
  J_i(t)=\frac12\bigl[A_i(t)+B_i(t)\bigr]
\label{eq:Ji_def}
\end{equation}
is the standard Ji angular-momentum combination; in the forward limit, $J_i(0)$ gives the total angular momentum carried by quarks or gluons in the gauge-invariant EMT decomposition~\cite{Ji1997}.  Equation~\eqref{eq:full_contraction_ABD} may therefore be written in terms of
\begin{equation}
3B_i(t)-D_i(t)=6J_i(t)-3A_i(t)-D_i(t).
\label{eq:AJD_combo}
\end{equation}
This rewriting is not only algebraic: it expresses the off-forward chromoelectric spin-two contribution through momentum, angular-momentum, and mechanical GFFs.

After dividing out the common scalar spinor factor, the finite-$t$ ratio is
\begin{equation}
R_{2/0}^{\rm full}(t)=
\frac{\displaystyle
\sum_{i=q,g}C_i^{(2)}\left[
\frac34MA_i(t)+\frac{t}{16M}\left(3B_i(t)-D_i(t)\right)
\right]}
{\displaystyle C_{\calA}\calA_N(t)+C_\sigma\sigma_N(t)} .
\label{eq:R_full_t}
\end{equation}
At $t=0$, the $B_i$ and $D_i$ terms vanish and Eq.~\eqref{eq:R_full_t} reduces to Eq.~\eqref{eq:R_forward}.  At finite $t$, these terms can change the slope of the charmonium photoproduction amplitude and therefore the observed shape of $d\sigma/dt$.

\subsection{Transverse slope rather than static three-dimensional radius}

In the $\xi=0$ frame, $t=-\Delta_\perp^2$.  For a normalized form factor $F(t)$, a light-front transverse density may be defined by
\begin{equation}
  \rho_F(b_\perp)=\int\frac{d^2\Delta_\perp}{(2\pi)^2}
  e^{-i\Delta_\perp\cdot b_\perp}F(-\Delta_\perp^2),
\end{equation}
when the conditions for a density interpretation are satisfied.  The associated transverse mean-square scale is
\begin{equation}
  \langle b_\perp^2\rangle_F
  =4\left.\frac{d}{dt}\ln F(t)\right|_{t=0}.
\label{eq:LF_transverse_radius}
\end{equation}
This is not the same convention as the three-dimensional slope parameter $6\,d\ln F/dt|_{0}$.  In the phenomenological discussion below, slopes are interpreted as scalar or gravitational light-front transverse profile scales.  This avoids treating the measurement as an instantaneous three-dimensional determination of a proton boundary.

\section{Numerical benchmark and robustness}\label{sec:numerics}

The finite-$t$ analysis combines three types of input.  The forward normalizations $A_g(0)$ and $A_q(0)$ are fixed by the PDF moments in Table~\ref{tab:forward_pdf}.  The partonic GFF shapes $A_i(t)$, $J_i(t)$, and $D_i(t)$ are modeled using lattice-guided dipole benchmarks based on the flavor/gluon decomposition in the range $0\le -t\le2~\GeV^2$~\cite{HackettPefkouShanahan2024}.  The scalar denominator is assigned a dispersive-guided scalar trace profile, with the radius varied in the robustness scan~\cite{CaoGuoLiYao2025}.  This setup is designed to test how the chromoelectric observable responds to different scalar and gravitational transverse profiles.  The finite-$t$ curves below are benchmark profiles, not fits to photoproduction data.

For reproducibility, the central curves use dipoles
\begin{equation}
  F_i(t)=\frac{F_i(0)}{(1-t/\Lambda_i^2)^2}
\label{eq:dipole_inputs}
\end{equation}
for each GFF listed in Table~\ref{tab:finite_t_inputs}.  The $A_i(0)$ entries are the CT18NNLO PDF moments at $Q_{\rm LF}=2~\GeV$.  The $J_i(0)$ entries are the lattice-guided values rescaled so that $J_q(0)+J_g(0)=1/2$, and the default $D_i(0)$ entries are the lattice-guided central values.  The scalar denominator is multiplied by a normalized dipole with the scalar dipole slope parameter quoted as $r_s=0.97~\fms$, $F_s(t)=(1-t/\Lambda_s^2)^{-2}$ and $\Lambda_s^2=12/r_s^2$ after converting $r_s$ to $\GeV^{-1}$.  The robustness scan varies $r_s=0.75,0.85,0.97,1.10~\fms$ and rescales the total $D$ term by $\eta_D=0,0.5,1,1.5$, and by $\eta_D=0.873$ to match the dispersive total $D(0)=-3.38$.

\begin{table}[!htbp]
\centering
\caption{Central finite-$t$ benchmark inputs.  The dipole masses $\Lambda$ are given in GeV.  The quark row denotes the singlet quark sector used in the spin-two numerator.}
\label{tab:finite_t_inputs}
\begin{tabular}{lcccccc}
\toprule
Sector & $A_i(0)$ & $\Lambda_A$ & $J_i(0)$ & $\Lambda_J$ & $D_i(0)$ & $\Lambda_D$\\
\midrule
$q$ & 0.5866 & 1.477 & 0.2480 & 1.620 & $-1.30$ & 0.811\\
$g$ & 0.4134 & 1.262 & 0.2520 & 1.399 & $-2.57$ & 0.538\\
\bottomrule
\end{tabular}
\end{table}

For the central benchmark, CT18NNLO at $\mu_h=2~\GeV$, scalar trace profile corresponding to $r_s=0.97~\mathrm{fm}$ in the conventional slope convention, and the default lattice-guided $D$ normalization, the ratio increases from
\begin{equation}
  R_{2/0}^{\rm full}(0)=0.147
\end{equation}
to
\begin{equation}
  R_{2/0}^{\rm full}(|t|=0.2)=0.213,
  \qquad
  R_{2/0}^{\rm full}(|t|=0.5)=0.303,
  \qquad
  R_{2/0}^{\rm full}(|t|=1.0)=0.405 .
\end{equation}
The full $A_i,J_i,D_i$ contraction reduces the pure-$A_i$ spin-two numerator by approximately $6.7\%$, $11.6\%$, and $19.6\%$ at $|t|=0.2$, $0.5$, and $1.0~\GeV^2$, respectively.  Thus the $D$-term does not affect the forward normalization, but it can noticeably modify the finite-$t$ numerator.

\begin{figure}[!htbp]
  \centering
  \includegraphics[width=0.82\linewidth]{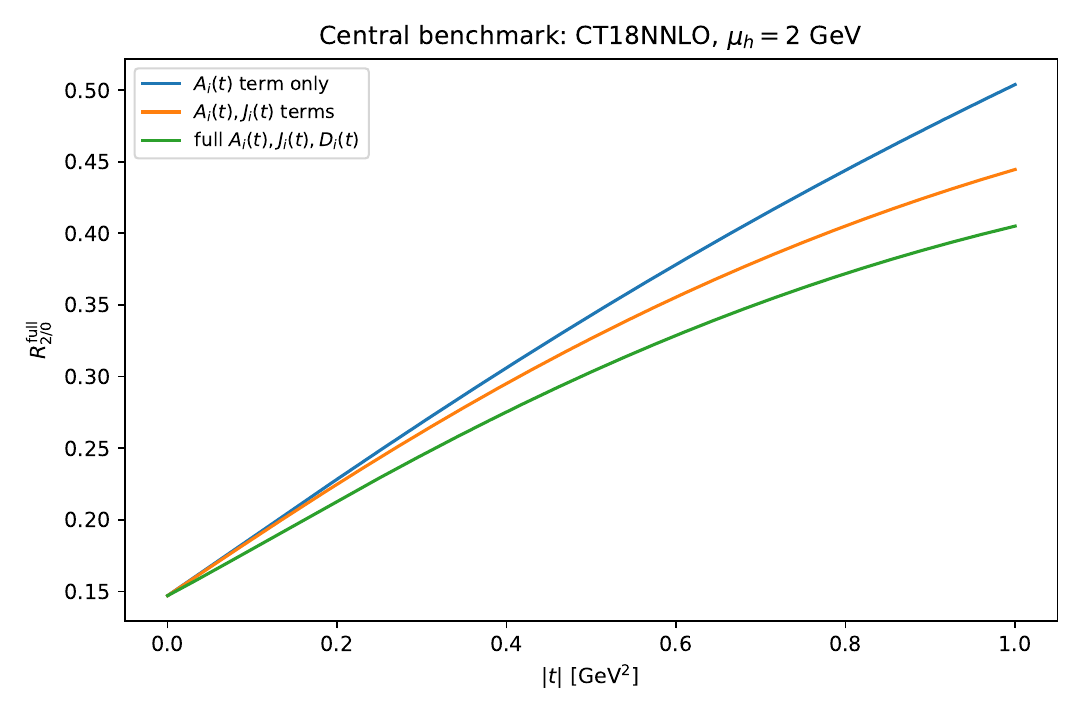}
  \caption{Central benchmark decomposition of $R_{2/0}^{\rm full}(|t|)$ for CT18NNLO and $\mu_h=2~\GeV$.  The three curves show the $A_i(t)$ term only, the $A_i(t),J_i(t)$ terms, and the full $A_i(t),J_i(t),D_i(t)$ contraction.  The caption notation follows Eq.~\eqref{eq:AJD_combo}.}
  \label{fig:central_decomp}
\end{figure}

Figure~\ref{fig:central_decomp} shows the central decomposition.  It makes the operator result of Eq.~\eqref{eq:R_full_t} visible numerically: the pure $A_i(t)$ curve represents the momentum-form-factor approximation, the $A_i(t),J_i(t)$ curve adds the angular-momentum part through $6J_i-3A_i$, and the full curve includes the mechanical $D_i(t)$ term.  The physical consequence is that the charmonium chromoelectric interaction is an operator filter rather than a direct projection onto one EMT component.  The scalar branch fixes the dominant forward normalization, while the spin-two branch carries a finite-$t$ sensitivity to momentum, angular momentum, and mechanical response.

\begin{figure}[!htbp]
  \centering
  \includegraphics[width=0.82\linewidth]{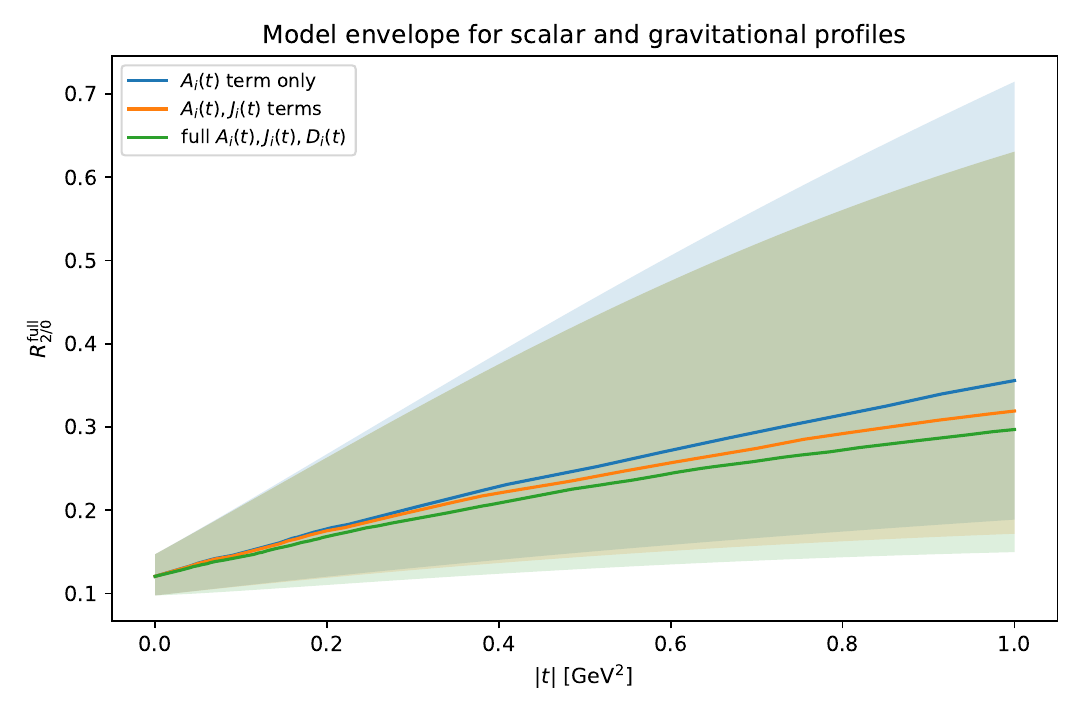}
  \caption{Robustness envelope for $R_{2/0}^{\rm full}(|t|)$ under variations of PDF set, matching scale, scalar transverse profile, and $D$-term normalization.  The central lines show the median behavior for the three spin-two truncations; the bands show the corresponding scanned envelopes.}
  \label{fig:R_envelope}
\end{figure}

Figure~\ref{fig:R_envelope} addresses a different question.  Once the central decomposition is understood, the relevant issue is whether the forward hierarchy is stable under realistic changes of PDF input, matching scale, scalar profile, and $D$-term normalization.  The figure shows that the forward $10$--$15\%$ spin-two fraction is robust, while the finite-$|t|$ behavior is governed by the relative scalar and gravitational slopes.  This makes the differential distribution a discriminator of form-factor profiles, rather than a single-number radius extraction.  The shaded bands are deterministic scan envelopes, not statistical confidence intervals.

The ratio $R_{2/0}^{\rm full}(t)$ measures the amplitude-level size of the spin-two contribution.  The experimentally fitted quantity is closer to the differential cross section, where scalar and spin-two terms interfere coherently.  I therefore also consider the distortion of the differential cross section relative to a scalar-only baseline.  This quantity asks whether the spin-two contribution would appear merely as a small correction to the amplitude or as a visible change in the measured $t$ distribution.

The reduced EMT matrix elements determine the relative scalar and spin-two strengths, but the physical photoproduction amplitude may carry process-dependent phases from the production mechanism, real-part corrections, and final-state dynamics.  I write
\begin{equation}
  \mathcal M(t)=N_0(t)+e^{i\phi}N_2(t),
\end{equation}
so that
\begin{equation}
  \mathcal R_\sigma(t)=\frac{|N_0(t)+e^{i\phi}N_2(t)|^2}{|N_0(t)|^2}
  =1+2\cos\phi\,R_{2/0}^{\rm full}(t)+\left[R_{2/0}^{\rm full}(t)\right]^2 .
\label{eq:sigma_distortion}
\end{equation}
The choices $\phi=0$ and $\phi=\pi$ are constructive and destructive interference benchmarks.  They are not assumptions about the unique physical phase; they show how strongly $d\sigma/dt$ could respond to the spin-two component for a given value of $R_{2/0}^{\rm full}(t)$.

\begin{figure}[!htbp]
  \centering
  \includegraphics[width=0.82\linewidth]{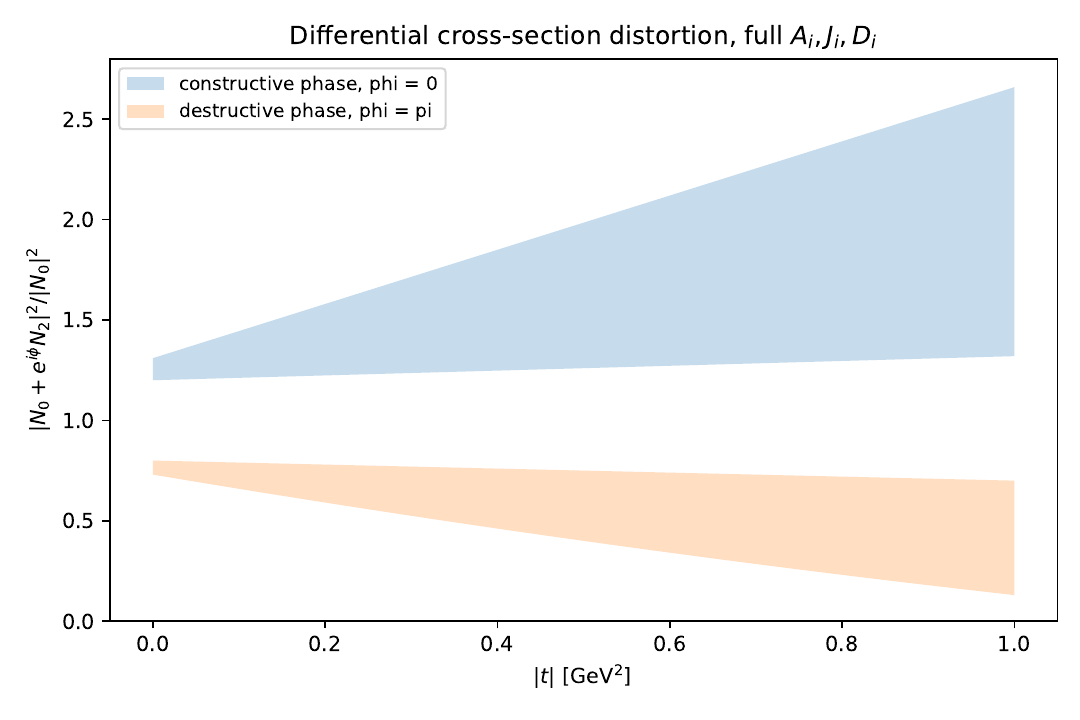}
  \caption{Envelope for the differential cross-section distortion relative to a scalar-only amplitude, using the full $A_i(t),J_i(t),D_i(t)$ contraction.  The two bands correspond to constructive ($\phi=0$) and destructive ($\phi=\pi$) scalar/spin-two interference benchmarks in Eq.~\eqref{eq:sigma_distortion}.}
  \label{fig:sigma_envelope}
\end{figure}

\FloatBarrier
\section{Discussion and conclusions}\label{sec:conclusions}

The chromoelectric operator selected by compact charmonium separates naturally into a scalar EMT contribution and a traceless spin-two contribution.  The scalar contribution is controlled by the anomaly/sigma combination $\calA_N=M-\Sigma_N$ together with the sigma term.  The forward light-front spin-two part is controlled by PDF momentum fractions, and its size is a subleading but meaningful correction to the scalar amplitude.  With modern PDFs, the forward ratio is about $0.10$--$0.15$ over the matching-scale range considered.

The off-forward result is more structurally informative.  The full chromoelectric spin-two contraction is not determined by $A_i(t)$ alone.  It contains
\begin{equation}
  3B_i(t)-D_i(t)=6J_i(t)-3A_i(t)-D_i(t),
\end{equation}
which connects the charmonium probe to angular-momentum and mechanical GFFs.  This gives a concrete route by which near-threshold charmonium data can test scalar and spin-two EMT profiles, rather than merely assigning a single mass radius to the proton.

In lattice- and dispersive-guided benchmarks, $R_{2/0}^{\rm full}(t)$ can grow at moderate $|t|$ because the scalar denominator falls faster than the spin-two numerator.  The robustness scan shows that this growth is best interpreted as a discriminator among form-factor profiles.  The light-front interpretation is essential here: the measured slope is most cleanly treated as a transverse gravitational or scalar profile scale, and only with additional assumptions as a spatial radius.

The main conclusion is that near-threshold charmonium observables probe a chromoelectric operator separating a dominant scalar anomaly/sigma amplitude from a subleading, but $t$-sensitive, spin-two gravitational amplitude.  Differential measurements can test this by comparing scalar and spin-two slopes and constraining $6J_i-3A_i-D_i$.

\appendix
\section*{Appendices}
\addcontentsline{toc}{section}{Appendices}

\section{Conventions and normalization checks}\label{app:conventions}

The conventions used throughout the manuscript are listed in Table~\ref{tab:conventions}.  The most important signs are the definition $\Delta=P'-P$, the spacelike convention $t=\Delta^2<0$, and the Dirac tensor $\sigma^{\mu\nu}=\frac{i}{2}[\gamma^\mu,\gamma^\nu]$.

\begin{table}[!htbp]
\centering
\caption{Conventions used in the calculation.}
\label{tab:conventions}
\begin{tabular}{ll}
\toprule
Object & Convention\\
\midrule
Metric & $g^{\mu\nu}=\mathrm{diag}(1,-1,-1,-1)$\\
Momentum transfer & $\Delta=P'-P$\\
Invariant transfer & $t=\Delta^2<0$ for spacelike transfer\\
Average momentum & $\bar P=(P'+P)/2$\\
On-shell identity & $\bar P^2=M^2-t/4$\\
Light-front coordinates & $x^\pm=(x^0\pm x^3)/\sqrt2$\\
Light-front metric & $g^{+-}=1$, $g^{++}=g^{--}=0$\\
Dirac tensor & $\sigma^{\mu\nu}=\frac{i}{2}[\gamma^\mu,\gamma^\nu]$\\
Traceless EMT & $O_i^{\mu\nu}=T_i^{\mu\nu}-\frac14g^{\mu\nu}\theta_i$\\
GFF angular momentum & $J_i(t)=\frac12[A_i(t)+B_i(t)]$\\
$D$-term convention & $D_i(\Delta^\mu\Delta^\nu-g^{\mu\nu}\Delta^2)/(4M)$\\
\bottomrule
\end{tabular}
\end{table}

\section{Spin-two coefficient evolution}\label{app:rg}

The singlet spin-two coefficients satisfy
\begin{equation}
\mu\frac{d}{d\mu}\binom{c_q}{c_g}
=\as
\begin{pmatrix}
A&-A\\ -B&B
\end{pmatrix}
\binom{c_q}{c_g},
\qquad
A=\frac{16}{3}C_F,
\qquad B=\frac{4}{3}n_f.
\end{equation}
For a gluonic matching condition at $\mu_h$, the solution can be written as
\begin{align}
 C_q^{(2)}(\mu)&=g_h^2\frac{A}{A+B}\left(1-\rho\right),\\
 C_g^{(2)}(\mu)&=g_h^2\frac{A+B\rho}{A+B},
\end{align}
with
\begin{equation}
\rho(\mu,\mu_h)=\left[\frac{\as(\mu)}{\as(\mu_h)}\right]^{-(A+B)/(2\beta_0)}.
\end{equation}
This evolution is the source of the matching-scale dependence in Table~\ref{tab:scale_scan}.

\section{Derivation of the off-forward contraction}\label{app:offforward_deriv}

Starting from Eq.~\eqref{eq:EMT_GFF}, contraction with $v_\mu v_\nu$ gives
\begin{align}
 v_\mu v_\nu\langle P'|T_i^{\mu\nu}|P\rangle
=\bar u(P')\bigg[&A_i(v\cdot\bar P)\slashed v
+B_i\frac{(v\cdot\bar P)i\sigma^{v\Delta}}{2M}
\nonumber\\
&+D_i\frac{(v\cdot\Delta)^2-v^2t}{4M}
+Mv^2\bar C_i\bigg]u(P).
\end{align}
The trace is
\begin{equation}
\langle P'|\theta_i|P\rangle
=\bar u(P')\left[A_i\slashed{\bar P}
+B_i\frac{i\sigma^{\bar P\Delta}}{2M}
-\frac{3t}{4M}D_i+4M\bar C_i\right]u(P).
\end{equation}
Forming $O_i^{\mu\nu}=T_i^{\mu\nu}-g^{\mu\nu}\theta_i/4$ cancels the $\bar C_i$ term.  With $v=\bar P/\sqrt{\bar P^2}$, one has $v^2=1$, $v\cdot\Delta=0$, and $v\cdot\bar P=\sqrt{\bar P^2}$.  The contraction reduces to
\begin{equation}
 v_\mu v_\nu\langle P'|O_i^{\mu\nu}|P\rangle
=\bar u(P')\left[\frac34A_i\slashed{\bar P}
+\frac{3B_i}{8M}i\sigma^{\bar P\Delta}
-\frac{t}{16M}D_i\right]u(P).
\end{equation}
The two on-shell identities
\begin{equation}
\bar u(P')\slashed{\bar P}u(P)=M\bar u(P')u(P),
\qquad
\bar u(P')i\sigma^{\bar P\Delta}u(P)=\frac{t}{2}\bar u(P')u(P)
\end{equation}
then give Eq.~\eqref{eq:full_contraction_ABD}.

\section[Dimensional origin of the chromoelectric matching]{Dimensional origin of Eq.~\eqref{eq:OE_match}}\label{app:matching}

This appendix records the short matching argument behind Eq.~\eqref{eq:OE_match}.  The point of the derivation is to keep the trace subtraction in $d=4-2\epsilon$ dimensions until after renormalization.  The finite scalar term proportional to $\gamma_m S$ is then fixed by the product of the evanescent part of the projector with the scalar-operator counterterm.

With a time-like vector $v^\mu$, define
\begin{equation}
  \mathcal O_{E,B}(v)=g_B^2\mathbf E_B^2(v)
  =-g_B^2\left(F_B^{a\mu\alpha}v_\alpha\right)
          \left(F^{a}_{B\,\mu\beta}v^\beta\right).
\end{equation}
The exactly traceless gluon spin-two operator in $d$ dimensions is
\begin{equation}
  O_{g,B}^{\mu\nu}
  =-F_B^{a\mu\alpha}F^{a\nu}_{B\ \ \alpha}
   +\frac{1}{d}g^{\mu\nu}F_B^2,
  \qquad g_{\mu\nu}O_{g,B}^{\mu\nu}=0 .
\end{equation}
It follows directly that the bare chromoelectric operator satisfies the $d$-dimensional identity
\begin{equation}
  \mathcal O_{E,B}(v)
  =g_B^2\left[v_\mu v_\nu O_{g,B}^{\mu\nu}
  -\frac{v^2}{d}F_B^2\right].
\label{eq:bare_d_identity}
\end{equation}
This is the place where the evanescent information enters: $1/d=1/4+\epsilon/8+O(\epsilon^2)$.

For the scalar part one needs the one-loop minimal-subtraction relation
\begin{equation}
  F_B^2=
  \left(1+\frac{\beta_0\as}{\epsilon}\right)[F^2]_R
  -\frac{12C_F\as}{\epsilon}S+O(\as^2),
\label{eq:F2_bare_scalar_mix}
\end{equation}
where $S=\sum_qm_q\bar q q$ is RG invariant at this order, and
\begin{equation}
  g_B^2=\mu^{2\epsilon}g^2
  \left(1-\frac{\beta_0\as}{\epsilon}\right)+O(\as^2).
\end{equation}
The $F^2$ poles cancel between coupling renormalization and Eq.~\eqref{eq:F2_bare_scalar_mix}, leaving the finite scalar coefficient $-g^2v^2[F^2]_R/4$ after minimal subtraction.  The sigma counterterm gives
\begin{align}
 -\frac{g_B^2v^2}{d}F_B^2
 &\supset
  g^2v^2\left(\frac14+\frac{\epsilon}{8}\right)
  \frac{12C_F\as}{\epsilon}S  \\
 &=g^2v^2\left[\frac{3C_F\as}{\epsilon}
   +\frac{3}{2}C_F\as\right]S .
\end{align}
After the pole is subtracted, the finite remnant is
\begin{equation}
  \frac{3}{2}C_F\as\,g^2v^2 S
  =\frac{g^2v^2}{4}\gamma_m S,
  \qquad \gamma_m=6C_F\as+O(\as^2).
\end{equation}
The spin-two counterterms generate only the standard quark/gluon spin-two mixing; for a purely gluonic matching condition their finite local coefficients at the matching scale are $C_g^{(2)}(\mu_h)=g^2(\mu_h)$ and $C_q^{(2)}(\mu_h)=0$ at this order.  Combining the finite scalar terms with the spin-two part gives Eq.~\eqref{eq:OE_match}.  Thus the $\gamma_m S$ term in Eq.~\eqref{eq:OE_match} is not an added phenomenological assumption; it is the finite remnant required by the $d$-dimensional trace projector and the scalar renormalization of $F^2$.


\begin{thebibliography}{99}

\bibitem{Peskin1979I}
M.~E.~Peskin, ``Short Distance Analysis for Heavy Quark Systems. 1. Diagrammatics,'' Nucl. Phys. B \textbf{156}, 365--390 (1979), doi:10.1016/0550-3213(79)90199-8.

\bibitem{BhanotPeskin1979II}
G.~Bhanot and M.~E.~Peskin, ``Short Distance Analysis for Heavy Quark Systems. 2. Applications,'' Nucl. Phys. B \textbf{156}, 391--416 (1979), doi:10.1016/0550-3213(79)90200-1.

\bibitem{Kharzeev2021MassRadius}
D.~E.~Kharzeev, ``Mass radius of the proton,'' Phys. Rev. D \textbf{104}, 054015 (2021), doi:10.1103/PhysRevD.104.054015, arXiv:2102.00110.

\bibitem{Duran2023GluonicGFF}
B.~Duran, Z.-E.~Meziani, S.~Joosten \textit{et al.}, ``Determining the gluonic gravitational form factors of the proton,'' Nature \textbf{615}, 813--816 (2023), doi:10.1038/s41586-023-05730-4, arXiv:2207.05212.

\bibitem{Jpsi007_2026}
J/$\psi$-007 Collaboration, S.~Joosten, Z.-E.~Meziani \textit{et al.}, ``Near-Threshold $J/\psi\to\mu^+\mu^-$ Photoproduction and the Gluonic Gravitational Form Factors of the Proton,'' arXiv:2602.14416.

\bibitem{CLAS12_2026}
P.~Chatagnon \textit{et al.} (CLAS Collaboration), ``Measurement of the near-threshold $J/\psi$ photoproduction cross section with the CLAS12 experiment,'' arXiv:2602.22128.

\bibitem{GuoJiYuan2023}
Y.~Guo, X.~Ji, and F.~Yuan, ``Proton's gluon GPDs at large skewness and gravitational form factors from near threshold heavy quarkonium photo-production,'' arXiv:2308.13006.

\bibitem{GuoYuanZhao2025}
Y.~Guo, F.~Yuan, and W.~Zhao, ``Bayesian Inferring Nucleon's Gravitation Form Factors via Near-threshold $J/\psi$ Photoproduction,'' arXiv:2501.10532.

\bibitem{PentchevChudakov2024}
L.~Pentchev and E.~Chudakov, ``Rosenbluth separation of the $J/\psi$ near-threshold photoproduction - an access to the gluon gravitational form factors at high $t$,'' arXiv:2404.18776.

\bibitem{BurkertRMP2023}
V.~D.~Burkert, L.~Elouadrhiri, F.~X.~Girod, C.~Lorc\'e, P.~Schweitzer, and P.~E.~Shanahan, ``Colloquium: Gravitational form factors of the proton,'' Rev. Mod. Phys. \textbf{95}, 041002 (2023), doi:10.1103/RevModPhys.95.041002, arXiv:2303.08347.

\bibitem{PolyakovSchweitzer2018}
M.~V.~Polyakov and P.~Schweitzer, ``Forces inside hadrons: pressure, surface tension, mechanical radius, and all that,'' Int. J. Mod. Phys. A \textbf{33}, 1830025 (2018), doi:10.1142/S0217751X18300259, arXiv:1805.06596.

\bibitem{HattaRajanTanaka2018}
Y.~Hatta, A.~Rajan, and K.~Tanaka, ``Quark and gluon contributions to the QCD trace anomaly,'' JHEP \textbf{12}, 008 (2018), doi:10.1007/JHEP12(2018)008, arXiv:1810.05116.

\bibitem{Tanaka2019Trace}
K.~Tanaka, ``Three-loop formula for quark and gluon contributions to the QCD trace anomaly,'' JHEP \textbf{01}, 120 (2019), doi:10.1007/JHEP01(2019)120, arXiv:1811.07879.

\bibitem{MetzPasquiniRodini2020}
A.~Metz, B.~Pasquini, and S.~Rodini, ``Revisiting the proton mass decomposition,'' Phys. Rev. D \textbf{102}, 114042 (2020), doi:10.1103/PhysRevD.102.114042, arXiv:2006.11171.

\bibitem{LorceMetzPasquiniRodini2021}
C.~Lorc\'e, A.~Metz, B.~Pasquini, and S.~Rodini, ``Energy-momentum tensor in QCD: nucleon mass decomposition and mechanical equilibrium,'' JHEP \textbf{11}, 121 (2021), doi:10.1007/JHEP11(2021)121, arXiv:2109.11785.

\bibitem{Petrov2024Radii}
V.~A.~Petrov, ``On Nucleon Radii,'' arXiv:2408.16679.

\bibitem{Petrov2026Sizes}
V.~A.~Petrov, ``On Sizes of Hadrons,'' arXiv:2603.14352.

\bibitem{FreeseMiller2021}
A.~Freese and G.~A.~Miller, ``Forces within hadrons on the light front,'' Phys. Rev. D \textbf{103}, 094023 (2021), doi:10.1103/PhysRevD.103.094023, arXiv:2102.01683.

\bibitem{FreeseMiller2021Unified}
A.~Freese and G.~A.~Miller, ``Unified formalism for electromagnetic and gravitational probes: Densities,'' Phys. Rev. D \textbf{105}, 014003 (2022), doi:10.1103/PhysRevD.105.014003, arXiv:2108.03301.

\bibitem{ShifmanVainshteinZakharov1978}
M.~A.~Shifman, A.~I.~Vainshtein, and V.~I.~Zakharov, ``Remarks on Higgs-boson interactions with nucleons,'' Phys. Lett. B \textbf{78}, 443--446 (1978), doi:10.1016/0370-2693(78)90481-1.

\bibitem{Ji1997}
X.~Ji, ``Gauge-invariant decomposition of nucleon spin,'' Phys. Rev. Lett. \textbf{78}, 610--613 (1997), doi:10.1103/PhysRevLett.78.610, hep-ph/9603249.

\bibitem{HackettPefkouShanahan2024}
D.~C.~Hackett, D.~A.~Pefkou, and P.~E.~Shanahan, ``Gravitational form factors of the proton from lattice QCD,'' Phys. Rev. Lett. \textbf{132}, 251904 (2024), doi:10.1103/PhysRevLett.132.251904, arXiv:2310.08484.

\bibitem{CaoGuoLiYao2025}
X.-H.~Cao, F.-K.~Guo, Q.-Z.~Li, and D.-L.~Yao, ``Dispersive determination of nucleon gravitational form factors,'' Nature Commun. \textbf{16}, 6979 (2025), doi:10.1038/s41467-025-62278-9, arXiv:2411.13398.

\end{thebibliography}
\end{document}